\documentclass[english,showpacs,twocolumn]{revtex4-1}

\usepackage{mathptmx}
\usepackage[T1]{fontenc}
\usepackage[latin9]{inputenc}
\usepackage{color}
\usepackage{amsmath}
\usepackage{amssymb}
\usepackage{hyperref}
\usepackage{graphics}
\usepackage{graphicx}
\usepackage{epsfig}
\usepackage{bm}
\usepackage{epstopdf}
\usepackage{mathtools}

\usepackage{xmpmulti}

\def\etal{{\em et al.}}

\def\beq{\begin{equation}}
\def\eeq{\end{equation}}

\def\reff#1{(\ref{#1})}

\def\rhoc{\rho_\mathrm{c}}
\def\rhoi{\rho_\mathrm{i}}

\def\Ni{N_\mathrm{i}}
\def\Ne{N_\mathrm{e}}

\def\Wcmcm{\mbox{\rm W/cm$^{2}$}}

\def\omegaMie{\omega_\mathrm{M}}
\def\omegaM{\omega_\mathrm{M}}

\def\lambdaM{\lambda_\mathrm{M}}

\definecolor{dgreen}{rgb}{0.0, 0.5, 0.0}
\definecolor{dblue}{rgb}{0.0, 0.0, 0.5}

\def\Rinit{R_0}
\def\Re{R_\mathrm{e}}
\def\Ri{R_\mathrm{i}}

\def\v0{v_0}
\def\I0{I_0}

\def\me{m_\mathrm{e}}
\def\qe{q_\mathrm{e}}
\usepackage{babel}

\begin{document}

\title{
Dynamical resonance shift and unification of resonances in short-pulse laser cluster interaction
}
\author{S. S. Mahalik and M. Kundu}
\affiliation{Institute for Plasma Research, HBNI, Bhat, Gandhinagar - 382 428, Gujarat, India}
\date{\today}
\begin{abstract}
Pronounced maximum absorption of laser light irradiating a rare-gas or metal cluster is widely expected during the linear resonance (LR) when 
Mie-plasma wavelength $\lambdaM$ of electrons equals the laser wavelength $\lambda$.  
On the contrary, by performing molecular dynamics (MD) simulations of an argon cluster irradiated by short 5-fs (fwhm) laser pulses it is revealed that, for a given laser pulse energy and a cluster, at each peak intensity there exists a $\lambda$ -- shifted from the expected $\lambdaM$ -- that corresponds to a {\em unified dynamical} LR at which evolution of the cluster happens through very 
efficient
unification of possible resonances in various stages, including (i) the LR in the initial time of plasma creation, (ii) the LR in the Coulomb expanding phase in the later time and (iii) anharmonic resonance in the marginally over-dense regime for a relatively longer pulse duration, leading to maximum laser absorption accompanied by maximum removal of electrons from cluster and also maximum allowed average charge states for the argon cluster. 
Increasing the laser intensity, the absorption maxima is found to shift
to a higher wavelength in the band of $\lambda\approx (1-1.5)\lambdaM$
than permanently staying at the expected $\lambdaM$. 
A naive rigid sphere model also corroborates the wavelength shift of the absorption peak as found in MD and un-equivocally proves that maximum laser absorption in a cluster happens at a shifted $\lambda$ in the marginally over-dense regime of $\lambda\approx (1-1.5)\lambdaM$ instead of $\lambdaM$ of LR. 
Present study may find importance for guiding an optimal condition laser-cluster interaction experiment in the short pulse regime. 
\end{abstract}
\maketitle
\section{Introduction}\label{sec1}
Cluster of atoms or molecules, possessing solid-like atom density but of tiny size of a few nano-meter, facilitates full penetration of laser fields (of wavelengths $\lambda>100$~nm) without attenuation. It results very efficient coupling of laser with a cluster leading to
energetic electrons \cite{Kumarappan2002,Ditmire_PRA57,Springate_PRA61,Springate_PRA68,Shao_PRL77,Chen_POP_9,Kumarappan2003}, x-rays \cite{Jha_2005,Jha_2006,Chen_PRL104,McPherson_Nature370}, ions \cite{Ditmire_PRL78,Ditmire_Nature386,Ditmire_PRL78_2732,Kumarappan_PRL87,Lezius,Fukuda, Kumarappan2001,Krishnamurthy,Kumarappan2002,Ditmire_PRA57} and neutrals \cite{Rajeev_Nature}.
As a cluster is illuminated by a laser, constituent atoms are ionized (called inner ionization) and a nano-plasma is formed. Subsequently, laser absorption by electrons and removal of those hot electrons from the transient cluster potential (called outer ionization) creates a local electrostatic field which, when added to the laser field, may create {\em even} higher charge states (called ionization ignition \cite{RosePetruck,Bauer2003,Ishikawa,Bauer2004}) those are forbidden by the laser alone. 

For a short laser pulse duration, background ions remain relatively frozen and Mie-plasma frequency $\omegaM$ (corresponding to wavelength $\lambdaM$) remains much above the laser frequency $\omega$ for $\lambda > 400$~nm. In this regime, an electron may pass through anharmonic resonance (AHR) when its dynamical frequency $\Omega[r(t)]$ in the anharmonic potential gradually decreases and finally meets the driving $\omega$ \citep{MKunduprl,MKundupra2006}. The role of AHR as a dominant collisionless process in laser cluster interaction is established by theory \cite{Mulserprl,Mulserpra,Kostyukov}, particle-in-cell (PIC) simulations \citep{MKunduprl,MKundupra2006,Taguchi_PRl,Antonsen,MKundupra2012} and recently by molecular dynamics (MD) simulation \cite{SagarPOP2016}. AHR also finds its place for laser absorption in over-dense structured targets~\cite{MalayPOP2016,MKopex2015}. 

Assuming collective motion of electrons as a harmonic oscillator, $\ddot{x} + \omegaM^2 x = - E_0 \cos(\omega t)$, which may be satisfied by those bound electrons in the bottom of the potential; one finds the excursion as $x = - E_0\cos(\omega t)/(\omegaM^2 - \omega^2)$ and average energy 
$\langle\varepsilon\rangle = \langle (\omegaM^2 x^2 + \dot{x}^2)/2 \rangle 
= E_0^2(\omegaM^2 + \omega^2)/4(\omegaM^2 - \omega^2)^2$. In this collective oscillation model, pronounced and maximum absorption of laser in a cluster-plasma is expected when the linear resonance (LR) condition $\omegaM=\omega$ (or $\lambdaM=\lambda$) is satisfied.
Such a LR can be met in different ways, when (i) $\omegaM$ rises towards $\omega$ during the cluster charging (from the neutral condition) in the initial time, (ii) $\omegaM$ drops towards $\omega$ during the Coulomb expanding (CE) phase, and (iii) $\omega$ approaches $\omegaM$ by varying $\lambda$ of a tunable laser, in the sense that, $\lambda$ is changed from one experiment to the other.
Amongst these possibilities, the LR in the CE phase [option (ii)] has been demonstrated in experiments \cite{Zweiback,Saalmann_JPB39,Fennel_RMP,Ditmire_PRA53,Doppner,Koller,Zamith} with longer pulses > 50~fs. Due to non-availability of tunable lasers of any desired $\lambda$, the option (iii) can not be explored experimentally (always) for a cluster and one has to rely on models or numerical simulations. However, in previous simulations by Petrov \etal \cite{Petrov2005_PRE} and in subsequent studies 
\cite{Petrov2005,Petrov2006,Petrov2007} on laser heated clusters with pulses > 75~fs, no enhancement in the laser absorption and cluster charging were found while passing through the LR by $\lambda$ variation. Therefore, role of above LR was denied and the controversy {\em still} persists.  
For longer wavelengths, typically, $\lambda>400$~nm, the $\omegaM$ rises so sharply (due to faster ionization) that the LR condition is met instantaneously {\em only} for an infinitesimal time compared to the laser period, thus making this LR [option (i)] in the early time 
{\em very inefficient} 
for which no serious attention is paid to it by experimentalists.

Here, we study laser-cluster interaction in the fascinating regime of 5-fs (fwhm) short laser pulses.
The complex interplay between laser absorption, cluster charging, Coulomb expansion and electrons' outer-ionization depending upon laser and cluster parameters inhibits to predict an optimal condition for maximum laser absorption.
Hardly, one thinks of unifying all possible options [(i)-(iii)] in a single experiment.
The goal here is to look for the possibility to {\em unify} all kinds of LR [options (i)-(iii)] and also AHR in this short pulse regime, {\em particularly} making above options (i) and (iii) very efficient in a single laser pulse, and find out the optimal regime of laser wavelengths in the entire regime of intensities $\sim 10^{15}\Wcmcm - 10^{18}\Wcmcm$ for an argon cluster so that maximum conversion of laser energy into charged particle energy is obtained. 

To achieve our goal, the interaction of 5-fs (fwhm) laser pulses (of fixed energy but different $\lambda = 100-800$~nm) with an argon cluster is studied by our three dimensional MD simulation \cite{SagarPOP2016}.
It is found that, for a given pulse energy and a cluster, at each peak intensity there exists a wavelength $\Lambda_d$ -- shifted from the expected wavelength of Mie-resonance $\lambdaM$ -- that corresponds to a {\em unified dynamical} LR (we call it UDLR) at which evolution of the argon cluster happens through very efficient unification of possible resonances in various stages, including (i) the LR in the initial time of plasma creation, (ii) the LR in the CE phase in the later time and (iii) AHR in the marginally over-dense regime [nearly satisfying $\Lambda_d\approx(1-1.5)\lambdaM$] for a relatively longer pulse duration leading to maximum laser absorption accompanied by maximum outer-ionization and also maximum charging for the argon cluster. 
At this $\Lambda_d$ (typically in the UV regime), respective LR is found to be more efficient not only in the early cycles of laser interaction [option (i)] than the LR in the CE phase [option (ii)] but also it becomes very efficient than a longer $\lambda>400$~nm, due to significant time elapsed by the system near the LR during the rise of $\omegaM$, thus making options (i) and (iii) more viable for short 5-fs (fwhm) pulse. The dynamical unification of all possible options (i)-(iii) of LR and AHR within a single 5-fs (fwhm) pulse at the shifted wavelength of UDLR leading to maximum absorption (as shown here) is also rarely possible with a longer pulse and a longer $\lambda>400$~nm. 

In the conventional notion of LR, the absorption maxima should occur at $\lambda=\lambdaM$ irrespective of the laser intensity. However, in the present work, as the laser peak intensity increases, the maxima in the absorption and outer-ionization are found to grow together but gradually shift towards higher wavelengths in the band of $\Lambda_d\approx (1-1.5)\lambdaM$ for the argon cluster instead of absorption peaking at the expected $\lambdaM$. It means that observed redshift of the absorption maxima $\Delta\Lambda_d=\Lambda_d-\lambdaM$ can be as large as 40-50\% of $\lambdaM$ 
depending upon the laser intensity, cluster size and atomic constituents of the cluster.
MD results for the wavelength shift of resonance absorption peak are also justified by a simple rigid sphere model (RSM) \cite{SagarPOP2016} of cluster. 
Thus MD and RSM un-equivocally prove that maximum laser absorption in a cluster happens at a shifted wavelength in the marginally over-dense regime of $\Lambda_d\approx (1-1.5) \lambdaM$ due to UDLR instead of $\lambda=\lambdaM$ of LR. 
Based on above findings, we envisage that the current work may serve as a guideline to perform an optimal condition experiment for maximum laser absorption in a cluster at different intensities in the short pulse regime. It also (possibly) explains why Petrov \etal \cite{Petrov2005_PRE,Petrov2005,Petrov2006,Petrov2007} could not find maximum absorption at the expected $\lambdaM$ and removes some of the controversies.

Organization of the paper is as follows. Section~\ref{sec2} illustrates laser absorption in an uniformly charged argon cluster by RSM for the basic understanding of the shifting of the absorption peak from $\lambda=\lambdaM$ of LR. Section~\ref{sec3} gives its further detail and possible unification of resonances (LR and AHR, i.e., UDLR) through various stages of evolution of argon cluster by rigorous MD simulations. Summary and conclusion are given in Sec.\ref{sec6}.
Atomic units (i.e., $\me = \vert -e \vert =1, 4\pi\epsilon_0 = 1, \hbar = 1$) are used here unless mentioned explicitly. 
\section{Wavelength shift of resonance absorption using rigid sphere model}\label{sec2}
Before studying the absorption of laser pulses in an argon cluster
by MD simulation in Sec.\ref{sec3}, here we study the 
same by a RSM where cluster of radius $\Rinit$ is assumed to be pre-ionized and consists of uniformly charged spheres of argon ions and electrons of equal radii $\Ri = \Re = \Rinit$. 
The motion of massive ion sphere is neglected for short laser pulse < 15 fs and also the laser magnetic field for intensities < $10^{18}\,\Wcmcm$ as considered here.

The equation of motion (EOM) of the electron sphere in a linearly polarized 
laser field $E_l(t)$ along $x$-direction reads \cite{SagarPOP2016}
\begin{equation} \label{eq:ofmotion}
%
{\ddot{\vec{r}}}+{\vec {r}}g(r)/r = \hat{x} (\qe/\me) E_l(t)/R_0 
\end{equation}
where $\vec{r} = \vec{x}/R_0$ and $r = \left|\vec {r}\right|$. 
The electrostatic restoring field $g(r)$ and 
corresponding potential $\phi(r)$ are respectively,
\begin{equation} \label{eq:restoringforce}
g(r) = \omegaM^2 \times \begin{cases}
r &\text{if $0\leq r \leq 1$}\\
{1}/{r^2} &\text{if $r \geq 1$}
\end{cases}
\end{equation}
\vskip -0.5cm
\begin{equation} \label{eq:potential}
\phi (r) = \omegaM^2 R_0^2 \times \begin{cases}
{3}/{2}-{r^2}/{2} &\text{if \, $0\leq r \leq 1$}\\
{1}/{r} &\text{if \, $r \geq 1$}.
\end{cases}
\end{equation}
As long as the excursion $r$ of the center of the electron sphere remains 
inside the ion sphere, it executes a harmonic motion with a constant 
eigen-frequency $\Omega[r(t)]=\omegaMie$, where $\Omega[r(t)]$ stands for the anharmonic frequency of electrons as introduced in Refs.\cite{MKunduprl,MKundupra2006}. After crossing the boundary of the ion sphere, it experiences the Coulomb force and its motion becomes anharmonic \cite{SagarPOP2016} with gradual decrease in $\Omega[r(t)]$ for increasing $r>1$. 
When the irradiating laser wavelength is such that $\Omega[r(t)] = \omegaMie = \omega$, then LR absorption happens for the electron sphere with its center at $r\leq 1$. Otherwise, AHR absorption happens in the over-dense cluster plasma when $\Omega[r(t)]  = \omega$ is met by the electron sphere at a location $r>1$. Thus, RSM can be used to understand both LR and AHR processes for laser absorption in a cluster. Details of AHR using different RSMs are given in Refs.\cite{SagarPOP2016,MKunduprl}.

Since the cluster size is much smaller than wavelengths $\lambda=100-800$~nm, the effect of propagation of light (directed in $z$) is disregarded. 
Laser vector potential is defined in the dipole approximation as $A(z,t) = A(t)\exp(-i 2\pi z/\lambda) \approx A(t) = (E_0/\omega)\sin ^2 (\omega t/2n) \cos (\omega t)$ 
for $0<t<n T$; where $n$ is the number of laser period $T$, 
$\tau = n T$ is the total pulse duration and $E_0=\sqrt{8\pi I_0/c}$ 
is the field strength for the peak intensity $I_0$. 
Defining ${E_l} (t)=-d{A} /dt$, one finds \cite{SagarPOP2016,MKundupra2012} 
\begin{equation} \label{eq:laserfield}
{E_l} (t) = (E_0/\omega) \begin{cases}
\sum_{i=1}^{3}c_i\omega_i\sin(\omega_i t) &\text{if \, $0 < t < nT$}\\
0 &\text{otherwise};
\end{cases}
\end{equation}
where $c_1=1/2, c_2=c_3= -1/4, \omega_1 = \omega, 
\omega_2 = (1+1/n)\omega$, and $\omega_3 = (1-1/n)\omega$. 
%

%
In the RSM we need to assign a uniform (an average) charge state $\overline{Z}$ to define $\omegaM$, where $\omegaMie=\sqrt{4\pi\rhoi/3}$, $\rhoi = 3 N\overline{Z}/4\pi R_0^3$ is the charge density of ion background and $N$ is the number of atoms in the cluster. For an arbitrary value of $\overline{Z}$ irrespective of the laser intensity $I_0$, the $\omegaM$ may be over-estimated or under-estimated. 
Therefore, in the RSM, $\overline{Z}$ must be at least 
the value of the corresponding optical field ionization (OFI) of argon atoms/ions for a given $I_0$. To obtain $\overline{Z}$, we first create a table of $Z$ vs $I_0(Z)$ obeying the critical field $E_0(Z) = I_p^2 (Z)/4Z$ of OFI \cite{MKunduPOP2008,MKunduPRA2007,Popruzhenko2008}, where  $I_p(Z)$ is the ionization potential (IP) corresponding to the integer charge state $Z$ of an argon atom/ion.
Then from the table look-up (or interpolation from the $Z$ vs $I_0(Z)$ OFI curve) we find a value of $Z$ (and call it average $\overline{Z}$) for the given intensity of $I_0(\overline{Z})$.
Thus $I_0(\overline{Z})$ is regarded as the OFI intensity corresponding to an average $\overline{Z}$ which may be non-integer. 
In the case of MD simulation (in Sec.\ref{sec3}), however, each atom/ion is self-consistently ionized to different integer $Z$ only, but the final average charge $\overline{Z}$ per atom/ion may not be an integer.

We consider an argon cluster of $\Rinit=2.91$\,nm and $N=1791$. It is illuminated by respective peak intensities $I_0(\overline{Z}) \approx 5\times 10^{15}, 10^{16}, 5\times 10^{16}, 10^{17}\,\Wcmcm$ of OFI yielding different $\overline{Z} \approx 2.61, 3.42, 6.2, 6.94$ from the OFI curve, and take different number of non-interacting electron spheres $\Ne = N \overline{Z} = 4675, 6125, 11105, 12429$ in the RSM, with centers of electron spheres placed uniformly inside the ion sphere. The center of each electron sphere mimics a real point size electron.
For $\overline{Z}=1$, one obtains an uniform charge density 
$\rhoi = 3 N\overline{Z}/4\pi R_0^3 \approx 2.7\times 10^{-3}$ a.u. and $\omegaMie=\sqrt{4\pi\rhoi/3} \approx 0.104$ a.u. At $\lambda = 800$~nm, it corresponds to an over-dense plasma of $\rhoi/\rhoc \approx 9.95$ and $\omegaM/\omega \approx 1.82$, where $\rhoc \approx 1.75\times 10^{27} m^{-3}$ is the critical density. With $\overline{Z}=1-8$, the LR wavelengths are found to be in the range of $\lambdaM \approx 440 - 156$~nm.

Above multi-electron systems are now simulated. Dynamics of an electron sphere and corresponding laser energy absorption are studied with laser pulses of respective $I_0(\overline{Z})$
by solving Eq.\reff{eq:ofmotion}. For a given $I_0$ and $\overline{Z}$, $\lambda$ is varied from $100$~nm to $800$~nm; while keeping the total pulse duration $\tau$ and the pulse energy density $\epsilon_p = \int_0^{\tau} I(t) dt$ constant. Throughout this work, we keep $\tau \approx 13.5$~fs (fwhm $\approx$ 5~fs) which corresponds to $n=5$ periods at $\lambda = 800$~nm. As $\lambda$ changes from 800~nm to 100~nm,
the number of cycles in a pulse (for a fixed $\tau$) increases, electron spheres oscillate more, the nano-plasma passes from the over-dense to the under-dense regime, and the dominant absorption process should exhibit its signature. 
%
\begin{figure}
\includegraphics[width=1.0\linewidth]{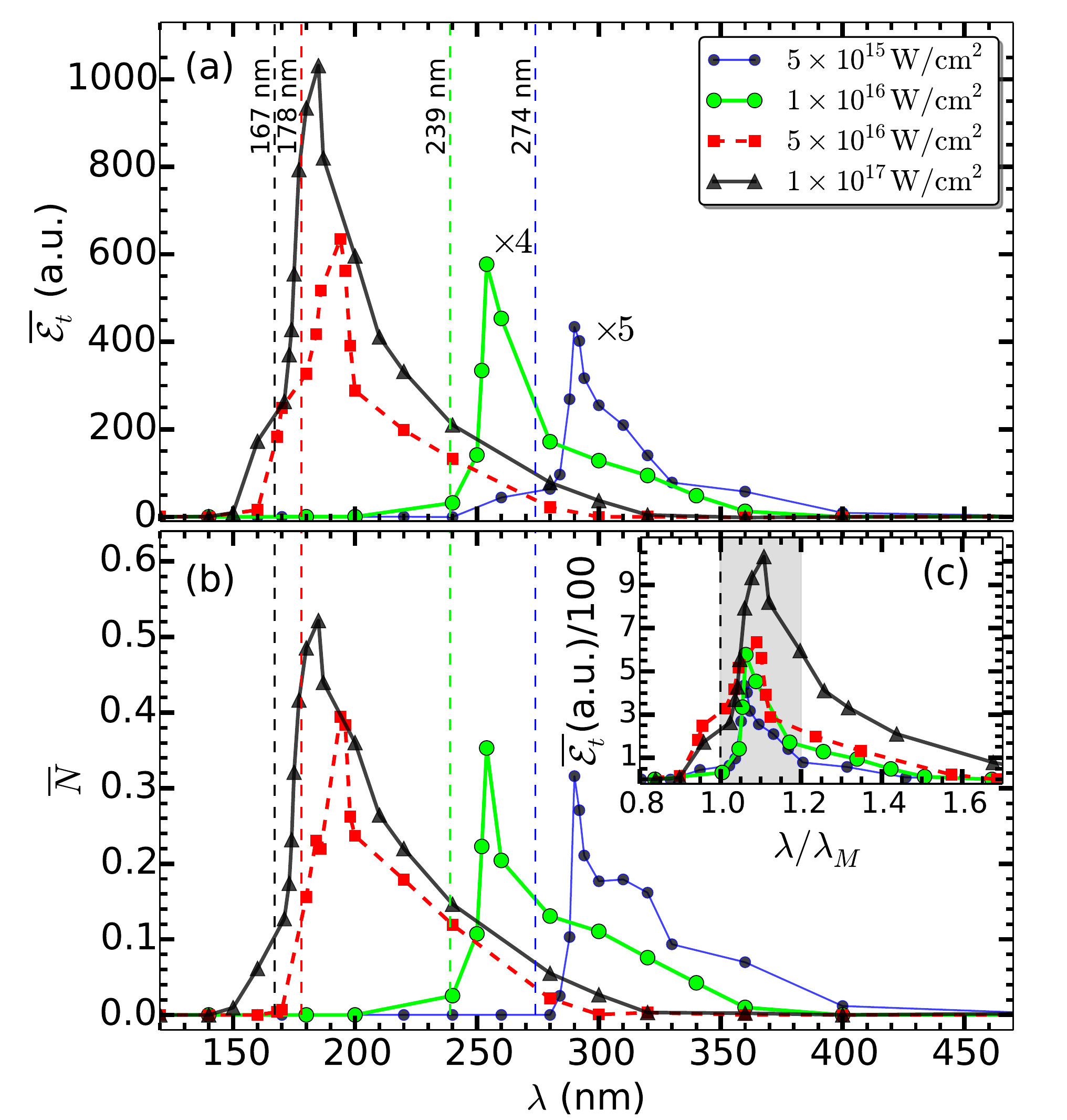}
\caption{(color online) Normalized total absorbed energy $\overline{\mathcal{E}}_t$ of electron spheres (a) and corresponding fractional outer ionization $\overline{N}$ (b) versus $\lambda$ for an argon cluster (radius $R_0=2.91$~nm, number of atoms $N=1791$ with various $\overline{Z} \approx 2.61, 3.42, 6.2, 6.94$) irradiated by laser pulses of respective OFI intensities $I_0(\overline{Z}) \approx 5\times 10^{15}, 10^{16}, 5\times 10^{16}, 10^{17}\,\Wcmcm$. For a given $I_0(\overline{Z})$, pulses of different $\lambda$ are chosen by keeping pulse duration $\tau \approx 13.5$~fs (fwhm $\approx$ 5 fs) as constant.
Vertical dashed lines indicate different $\lambdaM \approx 274, 239, 178, 167$~nm for respective $\overline{Z} = 2.61, 3.42, 6.2, 6.94$ where absorption maxima are strictly expected by LR. Inset (c) shows that absorption maxima are shifted in the marginally over-dense regime of $\lambda/\lambdaM\approx 1-1.2$. 
} 
\label{fig1abc}
\end{figure} 
%
\begin{figure}
\includegraphics[width=1.0\linewidth]{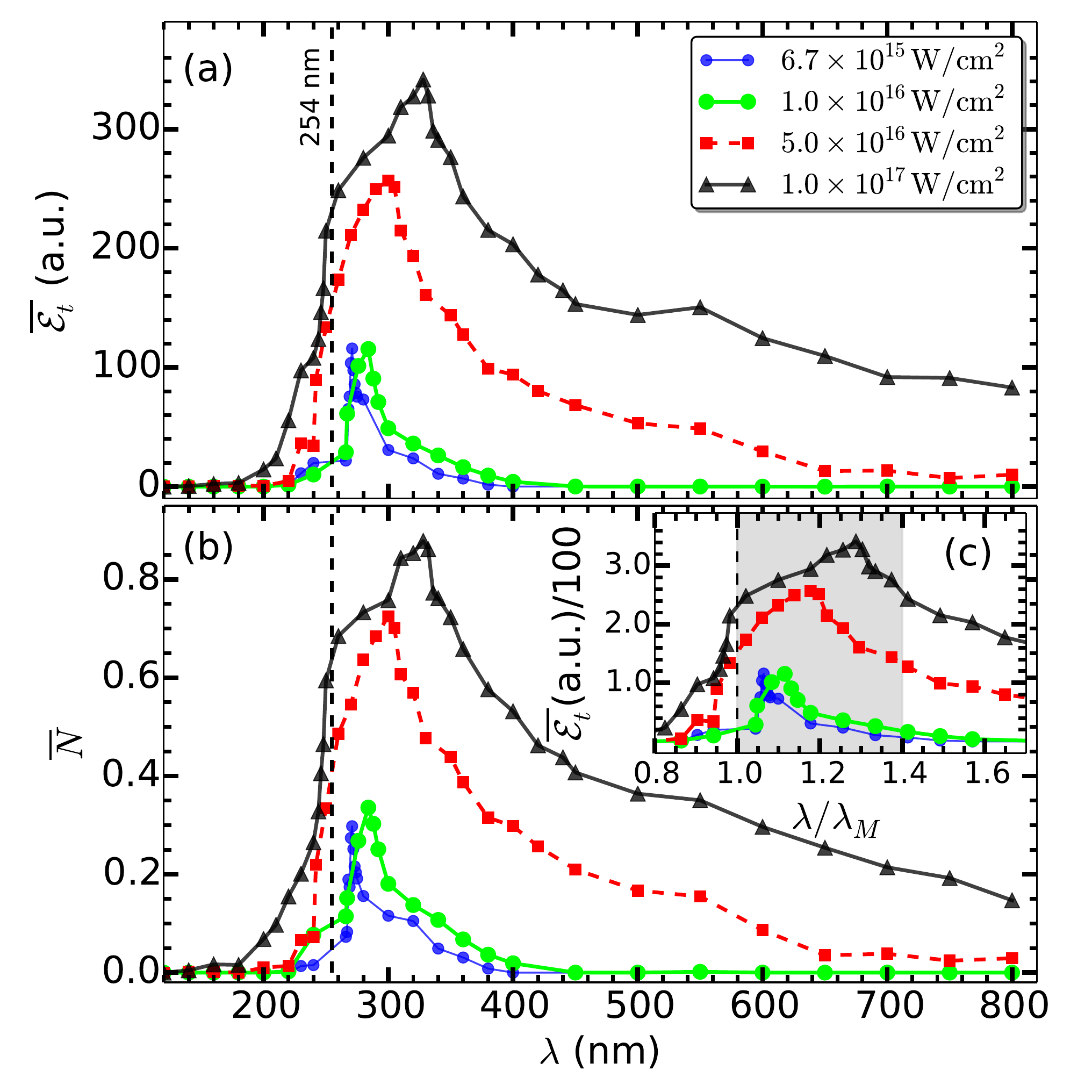}
\caption{(color online) Normalized total absorbed energy $\overline{\mathcal{E}}_t$ of electron spheres (a) and corresponding fractional outer ionization $\overline{N}$ (b) versus $\lambda$ for the argon cluster ($R_0=2.91$~nm, $N=1791$, with fixed $\overline{Z} = 3$ and $\Ne = N \overline{Z} = 5373$) irradiated by laser pulses with different $I_0$ starting from $I_0(\overline{Z}) \approx 6.7\times 10^{15}\,\Wcmcm$ of OFI for $\overline{Z}=3$. For a given $I_0$, pulses of different $\lambda$ are chosen by keeping pulse duration $\tau \approx 13.5$~fs (fwhm $\approx$ 5 fs) as constant. Vertical dashed line indicates corresponding $\lambdaM \approx 254$~nm where absorption maxima is strictly expected by LR. Inset (c) shows that absorption maxima are shifted in the marginally over-dense regime of $\lambda/\lambdaM\approx 1-1.4$. 
Other parameters are same as in Fig.\ref{fig1abc}.
} 
\label{fig2ab}
\end{figure} 
%

Figure~\ref{fig1abc}(a) shows the average total absorbed energy $\overline{\mathcal{E}}_t = \sum_1^{\Ne} (m_s v_i^2/2 + q_s\phi_i)/N$ by the electron spheres scaled by number of atoms $N$ (after subtracting the initial energy) at the end of the pulses (after $\tau=13.5$~fs) versus $\lambda$ for different $I_0 \le 10^{17}~\Wcmcm$ and $\overline{Z}$. 
Figure~\ref{fig1abc}(b) is the corresponding fraction of outer ionized electrons $\overline{N}=\Ne^{out}/\Ne$, where $\Ne^{out}$ is the total outer ionized electrons (out of $R_0$) after $\tau=13.5$~fs. 
Vertical dashed lines indicate different $\lambdaM \approx 274, 239, 178, 167$~nm (for respective $\overline{Z} \approx 2.61, 3.42, 6.2, 6.94$ above) where absorption maxima are strictly expected by LR.
For a combination of $I_0$ and $\overline{Z}$ (e.g., $I_0(\overline{Z}) \approx 5\times 10^{15}\,\Wcmcm$), it is found that $\overline{\mathcal{E}}_t$ and $\overline{N}$ increase with increasing $\lambda$, attain a maximum at a wavelength $\Lambda_d$ (e.g., $\Lambda_d\approx 290$~nm for $5\times 10^{15}\,\Wcmcm$), then drop as $\lambda$ is increased beyond $\Lambda_d$. Although the pulse energy is kept constant for all $\lambda$ for a given $I_0(\overline{Z})$, the occurrence of distinct maxima in the $\overline{\mathcal{E}}_t$ and $\overline{N}$ for all $I_0(\overline{Z})\lesssim 10^{17}\,\Wcmcm$ suggests the clear effect of $\lambda$. It also signifies more collective behavior of the electron system at $\Lambda_d$. However, absorption peaks are consistently found to be red-shifted from the respective $\lambdaM$ of LR for all $I_0(\overline{Z})$ which is a new finding by the RSM. Plotting 
$\overline{\mathcal{E}}_t$ against $\lambda/\lambdaM$ in Fig.\ref{fig1abc}(c) it is revealed that maxima in the absorption lie in the marginally over-dense regime of $\lambda/\lambdaM\approx 1-1.2$ for all OFI intensities.  

In the RSM, since ionic sphere charge density does not vary, the $\omegaM$ and $\lambdaM$ remain unchanged. We term the conventional condition $\lambdaM = \lambda$ (or $\omegaM=\omega$) as {\em static} LR and $\lambda=\lambdaM$ as the wavelength of static LR, to distinguish from the {\em unified dynamical} LR (we call it UDLR where LR and AHR are indistinguishable, more to be discussed later) in the marginally over-dense regime in presence of outer ionization (as already shown by the RSM), inner ionization and cluster expansion. 

Increasing the peak intensity for a fixed $\overline{Z}$ (e.g., for $\overline{Z} = 3$, this situation may arise when no more inner-ionization happens after the OFI) from the OFI intensity [e.g., $I_0(\overline{Z}=3)\approx 6.7\times{10^{15}}\,\Wcmcm$] as shown in Figs.[\ref{fig2ab}(a)-\ref{fig2ab}(b)], the maxima of $\overline{\mathcal{E}}_t$ and $\overline{N}$ increase in magnitude but gradually shift towards higher $\lambda$ from the expected static LR wavelength of $\lambdaM \approx 254$\,nm (vertical dashed line). Approaching towards $\lambdaM \approx 254$\,nm from 100~nm, higher intensity pulse expels more electrons from the cluster than at a lower intensity. When electrons move far from the cluster, the laser field dominates the restoring field of background ions acting on those free/quasi-free (outer-ionized) electrons. Since the average energy of a laser-driven free electron scales as $\approx E_0^2 \lambda ^2/4$ and electrons are liberated at different times with different energy, the redshift of absorption maxima from $\lambdaM$ depends on the population of free electrons during the interaction. As the free population of electrons increases with increasing intensity when $\overline{Z}$ is saturated, absorption peak gradually shifts towards higher $\lambda$ from the $\lambdaM$. 
Results presented in Fig.\ref{fig2ab} just represent a scenario for the redshift of the absorption peak if $I_0$ is increased when there is a saturation of $\overline{Z}$. {\textcolor{blue}{A similar situation has also been obtained for deuterium clusters (not shown here)}} with increasing $I_0$, where $\overline{Z}$ is quickly saturated at $\overline{Z}=1$ in beginning of the laser pulse and remaining part of the pulse is utilized for increasing outer ionization of electrons and gradual redshift of the absorption peak from the corresponding $\lambdaM$.
 Certainly, an under-estimated $\overline{Z}$ for an intensity will 
give more redshift than the actual. However, a better and a self-consistent estimation of $\overline{Z}$ at a given $I_0$ can only be obtained by MD simulations. 

At 400~nm wavelength, the argon cluster in Fig.\ref{fig2ab} is 
$\approx 7.46$ times over-dense with $\omegaM/\omega \approx 1.58$. For a low intensity $<10^{16}\,\Wcmcm$, almost all electrons remain bound in the harmonic part of the potential \reff{eq:potential} with nearly the same energy as the initial energy and negligible fractional outer ionization occurs after the pulse [see Figs.\ref{fig2ab}(a)-\ref{fig2ab}(b)]. Clearly, LR has no role beyond $\lambda = 400$~nm. {\em Still}, significant absorption and outer ionization persists for $I_0> 10^{16}\Wcmcm$. In this regime, as $I_0$ is increased, excursions of electron spheres increase beyond $r=1$, dynamical frequency $\Omega[r(t)]$ of each electron sphere drop from its initial value of $\omegaM$ and may meet the laser $\omega$, i.e., the AHR absorption condition $\Omega[r(t)] = \omega$ is met (see Refs.\cite{MKunduprl,MKundupra2006, SagarPOP2016}) for those electron spheres. In the intermediate regime of $\omegaM/\omega=\lambda/\lambdaM\approx 1-1.5$, effects of LR and AHR often become indistinguishable (the UDLR regime) with dominant contribution due to LR (and/or near LR fields) as evident from the occurrence of the absorption peak. Note that the approximate effective field $E_{eff}\approx \vert E_0/(\omegaM^2/\omega^2 -1)\vert$ inside the cluster is greatly 
 enhanced (both in the under-dense and over-dense regime, e.g., $1/2\lesssim (\lambda/\lambdaM)^2\lesssim 2$ ) near the LR and symmetric about $\lambda=\lambdaM$ where it has a peak. Whereas AHR can work {\em only} in the over-dense regime of $\lambda>\lambdaM$, absorption due to AHR monotonically decreases with increasing $\lambda>\lambdaM$ at a given laser intensity, and AHR can not solely produce an absorption peak without the dominant contribution of near LR enhanced fields. Here, near-the-LR augmented fields $E_{eff}>E_0$ (for $1\lesssim (\lambda/\lambdaM)^2\lesssim 2$) help some/many electrons to undergo AHR at ease. For higher intensity of $10^{17}\,\Wcmcm$, the absorption and outer ionization maxima are found at $\lambda=\Lambda_d\approx 330$~nm with $\lambda/\lambdaM\approx 1.3$ [see inset Fig.\ref{fig2ab}(c)]. In reality, however, at $10^{17}\,\Wcmcm$, charge states $\overline{Z}>3$ are possible by OFI, leading to increasing restoring force on electrons, reducing fractional outer ionization and possible shifting of the absorption peak somewhere in the marginally over-dense regime of $\Lambda_d \approx (1 - 1.5)\lambdaM$ as predicted in Fig.\ref{fig1abc}.     

From the simple RSM analysis we find that UDLR has a prominent role on maximum laser absorption and outer ionization. The absorption peak is justified to be red-shifted from the static LR wavelength $\lambdaM$ (in the absence of further inner-ionization), since there is always a fraction of free population of electrons with positive energy (for $I_0>10^{15}\,\Wcmcm$) which causes this shift. 
Therefore, {\em even if} $\omegaM$ remains fixed, the absorption maxima will {\em never} be found at the pre-calculated static LR wavelength of $\lambdaM$. Instead, it occurs at an un-predictable value in the shifted band of wavelengths $\Lambda_d\approx (1-1.5)\lambdaM$ that depends upon the level of outer ionization which in turn depends upon the laser intensity. In this shifted band of $\Lambda_d$, evolution of the cluster happens in the marginally over-dense regime -- the UDLR regime -- where AHR and influence of LR become indistinct and maximum absorption and outer ionization occur taking the benefit of the both. 

\section{Unified resonance absorption by MD simulation}\label{sec3}
Although RSM brings out basic features of laser-cluster interaction during the LR and AHR, many important aspects: (i) cluster expansion due to ion motion, (ii) cluster charging via polarization, and (iii) creation of enhanced charge states beyond OFI via ionization ignition can not be addressed by the RSM. Moreover, a fixed ion-potential is presumed in the RSM. In reality, potential should evolve in time, starting from zero, depending upon the spatial distribution of charges.
  
Previously, Petrov \etal \cite{Petrov2005_PRE,Petrov2006} performed MD simulations for a xenon cluster {\em only} at three laser wavelengths of 100~nm, 248 nm, and 800~nm at an intensity of $10^{16}\,\Wcmcm$ and concluded that, (i) cluster charging, (ii) average charge per atom, (iii) number of electrons, and (iv) peak electron density do not depend on laser wavelengths \cite{Petrov2006}. At $\lambda = 248$~nm of KrF laser, maximum laser absorption was expected due to the static LR.
However, they {\em failed} to find any enhancement in the absorbed energy at $\lambda=248$~nm compared to 100~nm and 800~nm; and {\em discarded} the role of LR. The ``null effect'' of LR was vaguely argued due to the ``non-uniform electron density'' in the cluster \cite{Petrov2005_PRE}. Note that electron density is always non-uniform in an ionized cluster and in its vicinity which can not explain non-existence of resonance peak at $\lambda=\lambdaM$. 
On the contrary, RSM results in Sec.\ref{sec2} clearly demonstrate an indispensable and combined role of LR and AHR for maximum absorption. The most important outcome of the RSM is that, the maximum absorption in a cluster {\em hardly} occurs at the static LR wavelength (as expected by Petrov \etal \cite{Petrov2005_PRE,Petrov2006} and almost all in this field). Instead, it occurs in a shifted wavelength band of $\Lambda_d\approx (1-1.5)\lambdaM$ due to outer ionization depending upon the laser intensity. In order to gain deeper insight for the absorption peak shift and to demonstrate the unified role of possible LR and AHR (the UDLR) on a strong footing, we perform more realistic three dimensional MD simulation \cite{SagarPOP2016} which is free from above mentioned short-comings of the RSM.

\subsection{Details of MD simulation}\label{sec3a}
%
Earlier version of our MD code \cite{SagarPOP2016} is now improved to include the self-consistent ionization of atoms/ions (field ionization), the motion of ions and the laser magnetic field. In many works, ion dynamics is often neglected assuming it to be important only after tens of femtoseconds. We find that ion dynamics is important {\em even} for the short 5-fs (fwhm) pulses consider here, since it determines the cluster expansion and the Mie-frequency $\omegaM$ which in turn determines LR and AHR. 
We perform simulations with various initial configurations of the neutral cluster, but results of average absorbed energy, outer ionization, and cluster charging are less sensitive to initial configurations for the chosen cluster and laser parameters in this work. Number of neutral atoms $N$ are put randomly inside the cluster radius $R_0=r_w N^{1/3}$, where $r_w\approx 0.24 $~nm is the Wigner-seitz radius for argon cluster. Choosing $N=1791$ we obtain $R_0 \approx 2.91$~nm. 
The thermal velocity assignment of neutral atoms in the cluster is done by Gaussian random distribution generated by the Box-Muller transformation and post processed to correlate with neutral atom positions in such a way that surface atoms have larger thermal velocities as compared to core atoms.
The velocity distribution relates the initial temperature of the neutral cluster which is taken as one-fifth of the room temperature $\approx 0.025$~eV for simplicity.

Ionization of atoms/ions is treated by ``over the barrier'' ionization (OBI) model. The laser intensity is appropriately chosen such that it ionizes all neutral atoms initially at the same time $t=t_i$ and produces first ionization state $Z=1$. 
The value of $t_i$ depends upon laser intensity and wavelength. At a fixed intensity, values of $t_i$ may have minor differences depending upon the phase of the pulses of different wavelengths near $t_i$. 
The critical laser field $E_c=\vert \vec{E}_l(t_i)\vert$ for this OFI can be obtained (as in Sec.\ref{sec2}) from the condition 
\begin{equation}\label{Bethe}
E_c = I_p^2 (Z)/4Z,
\end{equation} 
where  $I_p(Z)$ is the IP corresponding to $Z$. 
After the initial ionization by OFI, the nano-plasma is formed with equal number of electrons and ions 
$\Ne = \Ni = N = 1791$. As time advances, the laser field $\vec{E}_l(t>t_i)$ may displace 
few electrons along the laser polarization and space charge field $\vert\vec{E}_{sc}(\vec{r}_i,t)\vert$ is 
created at each ion position $\vec{r}_i$ due to the breaking of charge neutrality (may be at a microscopic 
level) which may exceed $\vert\vec{E}_l(t)\vert$.
$\vec{E}_{sc}(\vec{r}_i,t)$ at the ion position $\vec{r}_i$ is contributed by 
all electrons and all other ions at $\vec{r}\ne \vec{r}_i$. Thus, the effect of plasma environment is 
taken into account for the ionization of ions. 
The total instantaneous field $\vec{E} (\vec{r}_i,t) = \vec{E}_l(t) + \vec{E}_{sc}(\vec{r}_i,t) $, including the plasma field, may create even higher charge states $Z>1$ of ions (ionization ignition) by satisfying the condition
\begin{equation}\label{Bethecondition}
\vert \vec{E} (\vec{r}_i,t)\vert \geq I_p^2 (Z+1)/4(Z+1),
\end{equation} 
and the number of electrons increase dynamically.
The position and velocity of a newly born electron are assumed same as the parent ion conserving the momentum and energy. 
 For simplicity, we use standard $I_p(Z)$ of argon atom/ions as in Refs.\cite{RosePetruck,Petrov2005_PRE,Petrov2005,Petrov2006,MKunduPRA2007,MKundupra2012,Ishikawa}, whereas due to the additive effect of the plasma field $\vec{E}_{sc}({\vec{r}_i,t})$ higher charge states $Z$ are created (in this work) for cluster atoms/ions than predicted by OFI~\reff{Bethe} for isolated argon atoms/ions, if only laser field is considered. 
By sophisticated MD simulations, it has been shown that electronic levels of atoms in a cluster may differ from those of a single atom \cite{Fennel2007,Mathias2010}. The electronic energies in cluster-ions are also lowered by electronic screening and by the presence of other plasma ions. 
We have not yet considered these corrections in $I_p(Z)$, which will be included in future work. As a result, charge states of argon ions (as reported here) may be little under-estimated and corresponding redshifts of absorption peaks may be little over-estimated for different laser intensities. However, we have also simulated deuterium cluster (not reported here) where $Z=1$ is saturated quickly by OFI~\reff{Bethe} and above effects on $I_p(Z)$ due to plasma environment are relatively less, but redshifts of absorption peaks are found to be very prominent. 

The EOM of $i$-th charge particle in the laser field (propagating in $z$) with the electric field $E_l(t)$ polarized in $x$ and the magnetic field $B_l(t)$ along $y$ reads,
\begin{equation}\label{eom1}
\displaystyle{\frac{d\vec{p_i}}{dt} = \vec{F_i}(r_i,v_i,t)} +  q_i \lbrace E_l(t) \hat{x} + \vec{v_i}\times B_l(t) \hat{y} \rbrace, 
\end{equation}
where $\vec{F_{i}} = \sum\limits_{j=1, i\ne j}^{N_p}{q_i q_j}\vec{r_{ij}}/{{r_{ij}^3}}$ 
is the Coulomb force on $i$-th particle of charge $q_i$ due to all 
other $N_p-1$ particles each of charge $q_j$ and $N_p$ is the total number of particles including ions and electrons.
Usually, $B_l(t)\approx E_l(t)/c \ll 1$ for intensities 
$< 10^{18}\,\Wcmcm$. Note that $\vec{F_i}$ is singular for a small separation $r_{ij} \rightarrow 0$.
To mitigate this Coulomb singularity and to avoid steep increase of $\vec{F_i}$, for $r_{ij} \rightarrow 0$, an artificial smoothing parameter $r_0$ is customarily added with ${r_{ij}}$.
Here, $r_0$ is chosen equal to $r_w$, since other
values of $r_0\ne r_w$ do not produce correct 
frequency $\omegaM$ (see Ref.\cite{SagarPOP2016} for a detail argument). 
The modified Coulomb force on $i$-th particle and 
the corresponding potential at its location are respectively given by, 
\begin{equation}\label{MD_force}
{\vec F_{i}} = \sum_{j=1,i\ne j}^{N_p} \frac{q_i q_j {\vec{r_{ij}} } }{{(r_{ij}^2+r_0 ^2)^{3/2}}}, \,\,\,\,\,
\phi_{i} = \sum_{j=1, i\ne j}^{N_p} \frac{q_j}{{(r_{ij}^2+r_0 ^2)^{1/2}}}.
\end{equation}
This modification of ${\vec F_{i}}$ and $\phi_{i}$ allows a charge particle to pass through 
another charge particle in the same way as in the PIC simulation. Thus it helps to study collisionless absorption processes in plasmas, e.g., resonances. Equation \reff{eom1} is solved by the velocity verlet time integration scheme with a tiny time step $\Delta t = 0.01$ a.u. $\approx 0.24$~attoseconds, to resolve the highest $\omegaM$ corresponding to the highest $Z$. 
\vspace{-0.2cm}
\subsection{Dynamical resonance shift and unification of resonances}
\label{sec3b}
\begin{figure}
\includegraphics[width=1.0\linewidth]{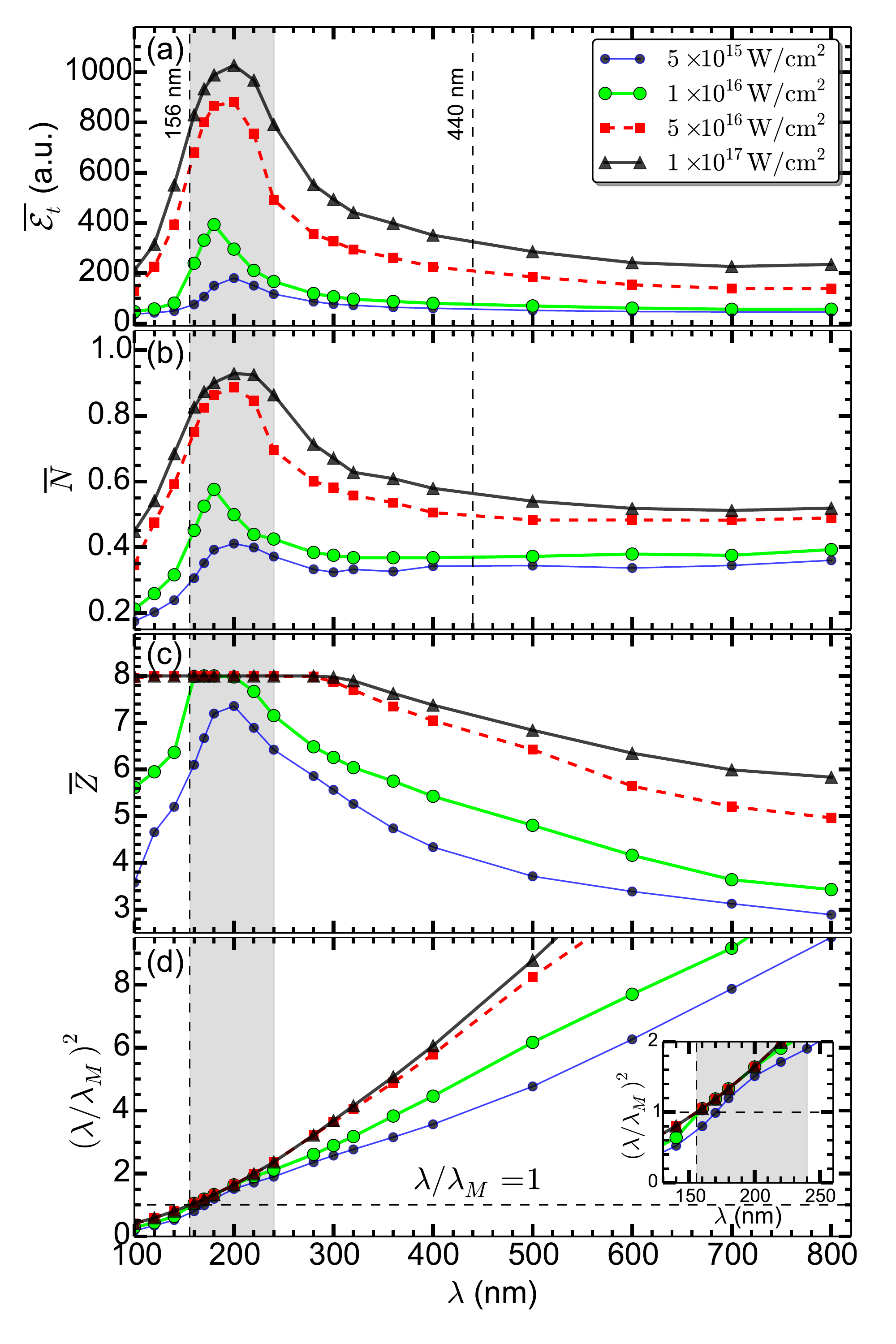}
\caption{(color online) 
Average total absorbed energy $\overline{\mathcal{E}}_t$ per atom~(a), corresponding fractional outer ionization $\overline{N}$ of electrons~(b), average  $\overline{Z}$ of argon ions~(c), and an approximate $(\lambda/\lambdaM)^2=(\omegaM/\omega)^2$~(d) versus $\lambda$ for the argon cluster (radius $R_0=2.91$~nm, number of atoms $N=1791$) after the laser pulses of peak intensities $I_0 \approx 5\times 10^{15}, 10^{16}, 5\times 10^{16}, 10^{17}\,\Wcmcm$ as in Fig.\ref{fig1abc}. For a given $I_0$, pulses of different $\lambda$ are chosen by keeping pulse duration $\tau \approx 13.5$~fs (fwhm $\approx$ 5 fs) as constant.
Vertical dashed lines indicate $\lambdaM$ where absorption maxima are strictly expected by LR for respective $\overline{Z} = 1,8$. Shaded bar highlights that absorption maxima are red-shifted in the marginally over-dense regime of $\lambda/\lambdaM\approx 1-1.5$, which is more clear in the inset of (d). 
}
\label{fig3abcd}
\end{figure}
The shift of the absorption peak as shown by the RSM becomes more un-predictable when there is cluster expansion and simultaneous creation of multiply charged ions during the laser interaction. 
For the above cluster of $N=1791$, $R_0 \approx 2.91$~nm; 
%
Figures~\ref{fig3abcd}(a)-\ref{fig3abcd}(d) show total absorbed energy $\overline{\mathcal{E}}_t = \sum_1^{N_p} (m_i v_i^2/2 + q_i\phi_i)/N$ per atom, fractional outer ionization $\overline{N} = \Ne^{out}/\Ne$ of electrons, average charge state $\overline{Z}=\Ne/N$ of ions (number of electrons produced $\Ne$ divided by number of atoms $N$) and an approximate $(\lambda/\lambdaM)^2=(\omegaM/\omega)^2$ versus $\lambda$ at the end of laser pulses (i.e., after 13.5 fs) of different peak intensities as in Fig.\ref{fig1abc} with RSM. 
Here $\lambdaM$ is calculated from the simple estimate of $\omegaM = ({N \overline{Z}/R_0^3})^{1/2}$ \cite{MKunduPRA2007,MKunduPOP2008,Popruzhenko2008}.
As in the RSM (Figs.\ref{fig1abc},\ref{fig2ab}), $\overline{\mathcal{E}}_t$ and the corresponding $\overline{N}$ for an intensity [Figs.\ref{fig3abcd}(a)-\ref{fig3abcd}(b)] increase with increasing $\lambda$, reach a maximum in the band of $\lambda\approx 196\pm 40$~nm and then decrease with further increase of $\lambda$. In addition, average $\overline{Z}$ also exhibit a peak [Fig.\ref{fig3abcd}(c)] in the above band of $\lambda$ at lower intensities $\le 10^{16}\,\Wcmcm$. For higher intensities $> 10^{16}\,\Wcmcm$, due to higher absorption and outer ionization even at lower wavelengths $<150$~nm, the average $\overline{Z}$ saturates at $\overline{Z}=8$ due to removal of all electrons from the $3s^2 3p^6$ shell of all argon atoms. Considering the OFI~\reff{Bethe}, only average $\overline{Z}\approx 2.61, 3.42, 6.2, 6.94$ are expected (as used in Fig.\ref{fig1abc}) at the respective peak intensities of $\approx 5\times 10^{15}, 10^{16}, 5\times 10^{16}, 10^{17}\,\Wcmcm$ for an isolated argon atom. However, for the {\em argon cluster}, the average $\overline{Z}$ exceeding the respective OFI values [Fig.\ref{fig3abcd}(c)] signifies the role of ``ionization ignition'' 
(by plasma space-charge fields $\vec{E}_{sc}$) which is found to be more efficient at lower intensities for the creation of higher $\overline{Z}$ than expected by OFI. As more electrons are removed from the $3s^2 3p^6$ shell, the ionization ignition becomes gradually weaker [an ``ionization depletion'' may also happen where $\overline{Z}$ falls below the OFI expected value \cite{MKunduPOP2008} for some $I_0$ and $\lambda$ as evident in Fig.\ref{fig3abcd}(c)] due to higher restoring force of ions on electrons. 
The field enhancement due to ionization ignition is not sufficient for the removal of next inner shell ($2s^22p^6$) electrons of argon atoms which require an intensity $> 10^{18}\Wcmcm$. 

For the argon cluster the static LR wavelength for $\overline{Z}=1$ is $\lambda_{M1}=\lambda_{M}\approx 440$~nm.
Since $\overline{Z}$ varies between $1-8$, the static LR wavelengths vary from $\lambda_{M1} \approx 440$~nm to $\lambda_{M8} \approx 440/\sqrt{8}\approx 156$~nm for $\overline{Z}=8$ [shown by vertical dashed lines in Figs.\ref{fig3abcd}(a)-\ref{fig3abcd}(b)]. 
However, the maxima in the absorption and outer ionization occur at shifted wavelengths (in spite of the saturation of inner-ionization at $\overline{Z}=8$) which lie in the band of $\lambda \approx 196\pm 40$~nm because of UDLR (i.e., combined dynamical LR and AHR) those are met differently for different peak intensities. 
Here, the dynamical LR, unlike the static LR, is decided self-consistently by the inner ionization, outer ionization and cluster expansion during the temporal evolution of a laser pulse and can not be decided {\em a priori}. At an intensity, the UDLR establishes an optimized competition between outer ionization, cluster expansion and inner ionization: the increasing outer ionization and cluster expansion momentarily try to push the absorption peak right-ward (not visible) from the static LR wavelength of $\lambda_{M1}=440$~nm in the initial time of OFI, while rapid creation of higher $\overline{Z}=2-8$ soon nullifies the expected red-shift and pushes the absorption peak opposite (ionization induced blue-shift towards $\lambda_{M8}\approx 156$~nm, left vertical dashed line) during the laser pulse driving. Finally, due to simultaneous outer ionization and cluster expansion, absorption peaks gradually shift right-ward in the band of wavelength $\Lambda_d\approx 196\pm40$~nm from the expected $\lambda_{M8}\approx 156$~nm with increasing intensity similar to the RSM.

Beyond $\lambda\approx240$~nm, argon cluster becomes significantly over-dense as seen from Fig.\ref{fig3abcd}(d) (and also from its inset) where $(\lambda/\lambdaM)^2 > 2$. In this regime of $\lambda>240$~nm, only AHR plays the dominant role \cite{SagarPOP2016,MKunduprl} as a collisionless process behind the absorption, outer ionization and charging of the argon cluster
and the approximate effective field $E_{eff}\approx E_0/(\omegaM^2/\omega^2 -1)$ in the cluster is suppressed below $E_0$. In the intermediate (marginally over-dense) regime of $\lambda\approx 156-240$~nm [i.e., $\lambda\approx (1-1.5)\lambdaM$, or $(\lambda/\lambdaM)^2\lesssim 1-2$, as seen from the inset of Fig.\ref{fig3abcd}(d)] where AHR and LR become indistinguishable (regime of UDLR), the LR and near LR enhanced field effects play the dominant role
as it is understood from the simple estimate of $E_{eff}\approx E_0/(\omegaM^2/\omega^2 -1)$, e.g., for $(\lambda/\lambdaM)^2 = 1.2, 1.5$, one obtains $E_{eff} = 5 E_0, 2 E_0$ respectively which may be little over-estimated form the reality. 
The near-LR enhanced field $E_{eff}\approx E_0/(\omegaM^2/\omega^2 -1)\gtrsim E_0$, in the marginally over-dense regime of $1\lesssim(\lambda/\lambdaM)^2\lesssim 2$, increases efficiency of AHR and helps some/many electrons to undergo AHR at ease.
Clearly, MD results with self-consistently determined charge states and cluster expansion (i.e., with variable $\omegaM$) in addition to the absorption and outer ionization as in the RSM, un-equivocally support that maximum absorption in a cluster happens in a shifted band of $\Lambda_d\approx (1-1.5)\lambdaM$, but not at the widely expected $\lambda=\lambdaM$ of static LR. 
{\textcolor{blue}{The relative red-shifts of absorption peaks are
$\Delta\Lambda_d/\lambdaM\approx 6.2\%, 7.5\%, 9.0\%, 9.6\%$ in Fig.\ref{fig1abc}, 
whereas in more realistic MD simulations (Fig.\ref{fig3abcd}) these are found to be 
$\Delta\Lambda_d/\lambdaM\approx 19.8\%,12.8\%,25\%, 28.2\%$ with increasing $\I0$}}.

Contrary to earlier findings of Petrov \etal \cite{Petrov2005_PRE,Petrov2006} for a laser driven xenon cluster, we find that 
cluster charging, average charge per atom, number of electrons, and peak electron density are strong function of $\lambda$ (see Fig.\ref{fig3abcd}) depending upon the peak laser intensity. Note that Petrov \etal \cite{Petrov2005_PRE,Petrov2006} performed MD simulations {\em only} at three different $\lambda =100, 248, 800$~nm at a single intensity of $10^{16}\,\Wcmcm$. Since the absorption maxima is red-shifted (as shown in this work by MD and RSM) in the regime of $\lambda\approx (1-1.5)\lambdaM$, it is possible that they had missed the absorption maxima which they precisely expected at $\lambda=\lambdaM=248$~nm (for the xenon cluster) according to the conventional notion.   

\begin{figure*}
\includegraphics[width=0.8\linewidth]{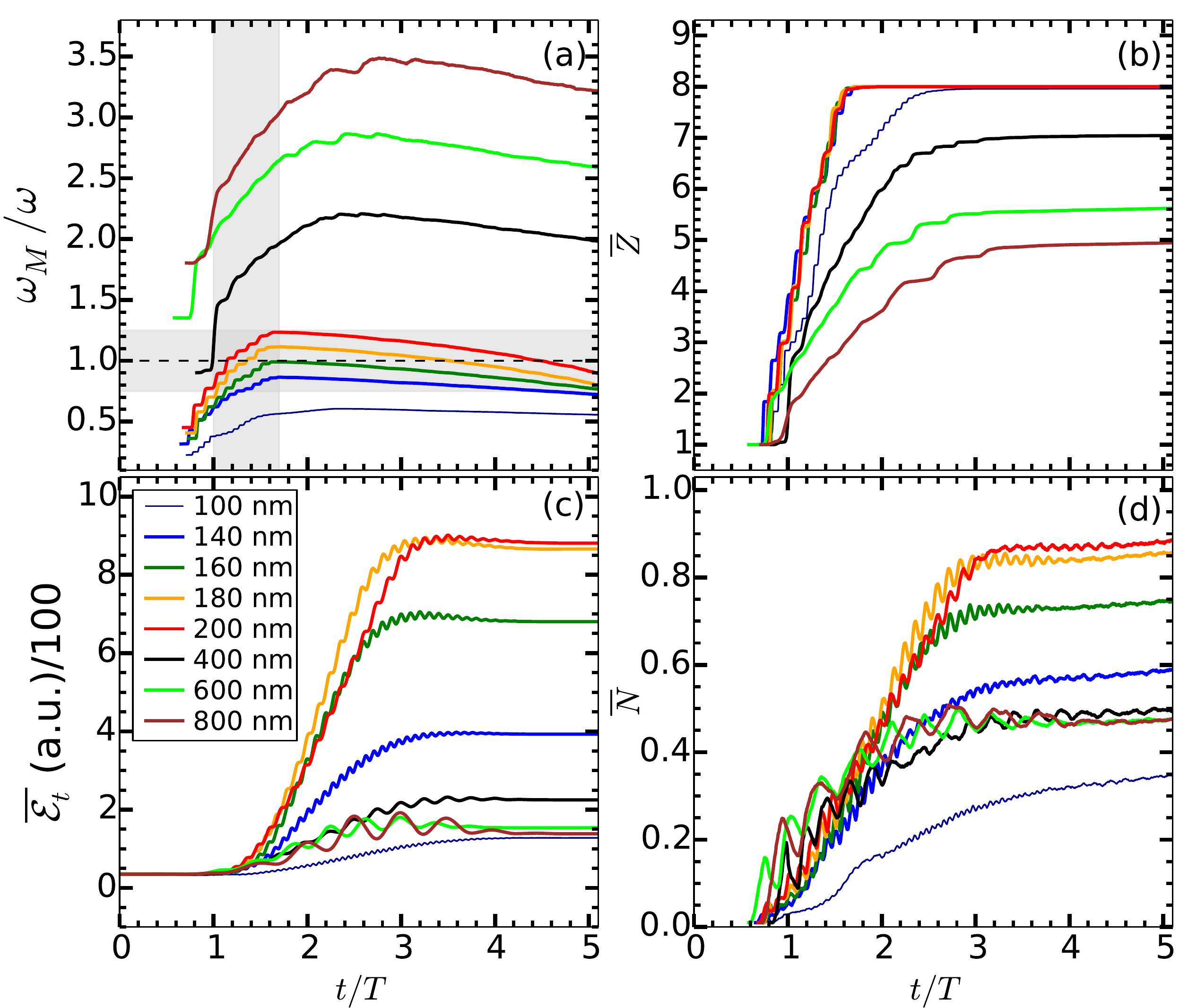}
\caption{(color online) Temporal variation of (a) normalized Mie-frequency $\omegaM(t)/\omega$, (b) average charge state $\overline{Z}(t)$, (c) total absorbed energy $\overline{{\mathcal{E}}}_t(t)$ per atom and (d) fraction of outer electrons $\overline{N}(t)$ for the argon cluster (in Fig.~\ref{fig3abcd}) when irradiated by 5-fs (fwhm) laser pulses of peak intensity $5 \times 10^{16}\,\Wcmcm$ and different $\lambda = 100-800$~nm. Time is normalized by the period $T$ of $\lambda=800$~nm. Other parameters are same as Fig.~\ref{fig3abcd}. It is seen that maximum absorption and outer ionization happen at the marginally over-dense $\lambda\approx 200$~nm. Shaded regions in (a) highlight efficient regime of interaction where UDLR is prominent.
}
\label{fig4abcd}
\end{figure*}

\subsubsection{Time domain analysis of the resonance shift}
\label{sec3bx}
To further justify the UDLR and the shifting of the absorption maxima, we analyze the evolution of the argon cluster in the time domain. Figures~\ref{fig4abcd}(a)-\ref{fig4abcd}(d) show temporal variation of scaled Mie-frequency $\omegaM(t)/\omega = \lambda/\lambdaM(t)$, average charge state ${\overline{Z}}(t)=\Ne(t)/N$ of an argon ion, total absorbed energy $\overline{\mathcal{E}}_t(t)$ per atom and fractional outer ionization $\overline{N}(t) = \Ne^{out}(t)/\Ne(t)$ of electrons respectively, for different $\lambda=100-800$~nm at the peak intensity of $5\times 10^{16}\,\Wcmcm$ corresponding to the result (red, dashed square) in Fig.~\ref{fig3abcd}. Time axis is normalized by the laser period $T$ of $\lambda =800$~nm. The horizontal dashed line in Fig.~\ref{fig4abcd}(a) indicates the line of static LR condition $\omegaM = \omega$. 
Since ion charge density is inhomogeneous due to variation of degree of ionization of ions at different locations at different times and due to the anisotropic cluster expansion \cite{MKunduPOP2008}; the exact calculation of dynamical $\omegaM(t)$ is difficult. To this end, we have tried the relation {\textcolor{blue}{$\omegaM^2(t) = Q_T(t)/R(t)^3$}}, where $Q_T(t)$ is the total ionic charge inside the dynamical cluster $R(t)$ which is the distance of the outermost ion in the expanding cluster at time $t$. 
Due to higher Coulomb expansion energy of outermost cluster ions than the ions in the cluster interior, the outer ion-layer dis-integrates much faster radially outward than the nearly homogeneously distributed, slowly moving interior ions. Therefore, when this $R(t)$ is taken into account, the dynamical $\omegaM(t)$ using above definition of $\omegaM(t)$ greatly under-estimates the actual $\omegaM(t)$ in the CE phase of the cluster and can not explain results in Fig.\ref{fig3abcd}. Hence, we calculate $\omegaM(t)$ from the relation {\textcolor{blue}{$\omegaM^2(t) = Q_0(t)/R_0^3$}} by assuming instantaneous total positive charge {\textcolor{blue}{$Q_0(t) = \Ni(t) \overline{Z}(t)$}} inside the initial cluster radius $R_0$ having $\Ni(t)$ number of ions with average charge $\overline{Z}(t)$ as in Refs.\cite{MKunduPRA2007,MKunduPOP2008,Popruzhenko2008}. {\textcolor{blue}{Excellent accuracy and compatibility of $\omegaM^2(t) = Q_0(t)/R_0^3$ are indeed confirmed {\em again} by time-frequency FFT analysis  (not reported here) of total dipole acceleration along the laser polarization as in Refs.\cite{MKunduPRA2007,Popruzhenko2008}.}}
The charging of the argon cluster starts around $t/T\approx 0.6-0.7$ for all wavelengths by the OFI~\reff{Bethe} when ${\overline{Z}}(t)=1$ and the dynamical $\omegaM(t)/\omega$ jumps from zero to a finite value [see Figs.~\ref{fig4abcd}(a)-\ref{fig4abcd}(b)]. Successively, laser absorption and outer ionization lead to ``ionization ignition'' which in turn creates higher $\overline{Z}(t)>1$ at different ion locations by meeting the condition \reff{Bethecondition} dynamically, thus causing $\overline{Z}(t)$ and corresponding $\omegaM(t)/\omega$ to increase in a step wise manner. 

For a shorter $\lambda <150$~nm, $\omegaM(t)/\omega$ remains below the line of static LR (under-dense) all the time, though respective $\overline{Z}(t)$ increases and saturates at $\overline{Z}(t) \approx 8$ before the peak of the laser pulse at $t/T=2.5$. In spite of a large number of inner ionized electrons $\Ne(t)\approx N \overline{Z}(t)$, absorption and outer ionization is low at 100~nm because of lower ponderomotive energy of electrons at 100~nm compared to 140~nm. Moreover, the $\omegaM(t)/\omega$ at 140~nm being more closer to the LR (from below), electrons experience more enhanced, near-the-LR effective field $E_{eff}\sim E_0/(\omegaM^2/\omega^2 -1)$ which is responsible for higher absorption and outer ionization than at 100~nm.

For a longer $\lambda > 300$~nm, the line of LR is passed too fast by $\omegaM(t)/\omega$ (see its sharp rise) just after the OFI in the initial time $t/T<1$ and $\omegaM(t)/\omega$ remains much above the line of LR (over-dense) all the time. This LR in the early time is not effective, due to weaker laser field and negligible time spent during (or near) the LR while crossing the resonance line. Just meeting the frequency matching condition of resonance is not enough for efficient transfer of energy from the driver to an oscillator. The time spent near the resonance and the strength of the driver are also important. Thus at long wavelengths > 300~nm, absorption and outer ionization are not so much in the early time $t/T < 1.5$. However, as time goes on, $\overline{Z}(t)$ increases and reaches a saturation while corresponding $\omegaM(t)/\omega$ reaches a maximum and drops due to cluster expansion after $t/T\approx 3$. The cluster charging, absorption and outer ionization for $\lambda>300$~nm are poorer than their respective values at the near-the-LR under-dense wavelength of $\lambda\approx 140$~nm. In this over-dense regime of $\lambda> 300$~nm,  clearly LR does not play {\em any role} but cluster charging, absorption and outer ionization still happen [Figs.\ref{fig4abcd}(b)-\ref{fig4abcd}(d)] due to the {\em dominant} AHR process \cite{MKunduprl,MKundupra2006,SagarPOP2016}.

In the intermediate band of $\lambda\approx 196 \pm 40$~nm, respective dynamical $\omegaM(t)/\omega$ just cross the line of static LR during $t/T\approx 1-1.6$ (vertical shaded bar) and elapse different times to its vicinity (in the marginally over-dense region, horizontal shaded bar) as compared to the other wavelengths outside this band. 
The slow passage of different $\omegaM(t)/\omega$ through the line of LR (i.e., more time elapsed near the LR) in the initial time $t/T<1.6$ couples laser energy so efficiently (in spite of laser field not being at its maximum) that inner-ionization is saturated due to the liberation of all electrons from $3s{^2}3p{^6}$ shells of argon atoms while absorption and outer ionization start at a greater pace as evident from change of their respective slopes.
At $160$~nm, which is close to the static LR condition of $\lambdaM\approx 156$~nm for $\overline{Z}\approx 8$, the $\omegaM(t)/\omega$ just meets the line of LR for a while $t/T\approx 1.5-2.0$, but absorption and outer ionization are {\em still} not at their maximum {\textcolor{blue}{which is often expected according to LR. Since $\omegaM(t)/\omega$ does not exceed unity, AHR is not possible}}. Instead, for 200~nm, the $\omegaM(t)/\omega$ slowly passes the line of LR during $t/T\approx 1-1.4$, reaches a maximum {\textcolor{blue}{value of $\omegaM/\omega\approx 1.25$ at $t/T\approx 1.6$}} and remains marginally above the line of LR for almost the entire pulse duration with a slow fall towards the under-dense regime due to Coulomb expansion. Here, resonances are efficiently and dynamically met at different stages, e.g., (i) LR during the initial time $t/T\approx 1-1.4$, (ii) combined LR and AHR in the marginally over-dense regime during $t/T\approx 1.2 - 4.5$, and (iii) LR in the Coulomb expansion phase around $t/T\approx 4.5$. The absorption and outer ionization are almost indistinguishable and remain almost equally efficient up to $t/T\lesssim 2.5$ for both 160~nm and 200~nm, but they 
begin to separate after the pulse peak at $t/T\approx 2.5$ where $\omegaM(t)/\omega$ quickly falls to the under-dense regime (where AHR is not possible) for 160~nm whereas it continues in the marginally over-dense regime (where both AHR and near-LR field effects contribute) for 200~nm until the end of the pulse. 
{\textcolor{blue}{The case of 180~nm is very similar to the case of 200~nm, i.e., passage of $\omegaM(t)/\omega$ through LR 
during $t/T\approx 1-1.2$, reaching a maximum value of $\omegaM(t)/\omega\approx 1.1$ at $t/T\approx 1.6$, traversal of $\omegaM(t)/\omega$ 
in the marginally over-dense regime during $t/T\approx 1-3.6$, meeting the LR in the CE phase at $t/T\approx 3.6$, and finally 
dropping to the under-dense regime for $t/T>3.6$. It is noticed that for 180~nm, $\overline{\mathcal{E}}_t(t)$ and $\overline{N}(t)$ grow in fater pace than 
those at 200~nm upto $t/T\approx 3$. However,}} 
the longer time spent by the system while passing through the LR and in the marginally over-dense regime justifies maximum absorption and outer ionization at the {\em shifted} wavelength of $\lambda = 200$~nm (from the static LR wavelength of $\lambdaM\approx 156$~nm) than 160~nm {\textcolor{blue}{and 180~nm as well}}.  
The average charge state, laser absorption and outer ionization of electrons [Figs.~\ref{fig4abcd}(b)-\ref{fig4abcd}(d)] are greatly enhanced at the end of the pulse of 200~nm due to the above unified dynamical LR which explains the maximum charging, absorption and outer ionization in Fig.\ref{fig3abcd} at different intensities as well as shifting of absorption peaks from the expected LR condition of $\lambda=\lambdaM$. 

Clearly, an effective unification of the early time LR [option(i)] along with the wavelength variation [option (iii)] in the band of $\Lambda_d \approx 196 \pm 40$, the LR in the Coulomb expanding phase [option(ii)] and the AHR in the marginally over-dense regime have been self-consistently decided by the system to work unitedly here (at 200~nm) for the maximum laser absorption in the argon cluster with a short 5-fs (fwhm) pulse which is rarely possible with long wavelength > 400~nm and longer pulses. Thus we find that unified dynamical LR (UDLR) leads to maximum laser absorption in a cluster in a shifted band of wavelength $\Lambda_d\approx (1-1.5)\lambdaM$ in presence of outer ionization.
%

\vspace{-0.6cm}
\subsubsection{Resonance shift with continuous short laser pulses}
\begin{figure}
\includegraphics[width=1.0\linewidth]{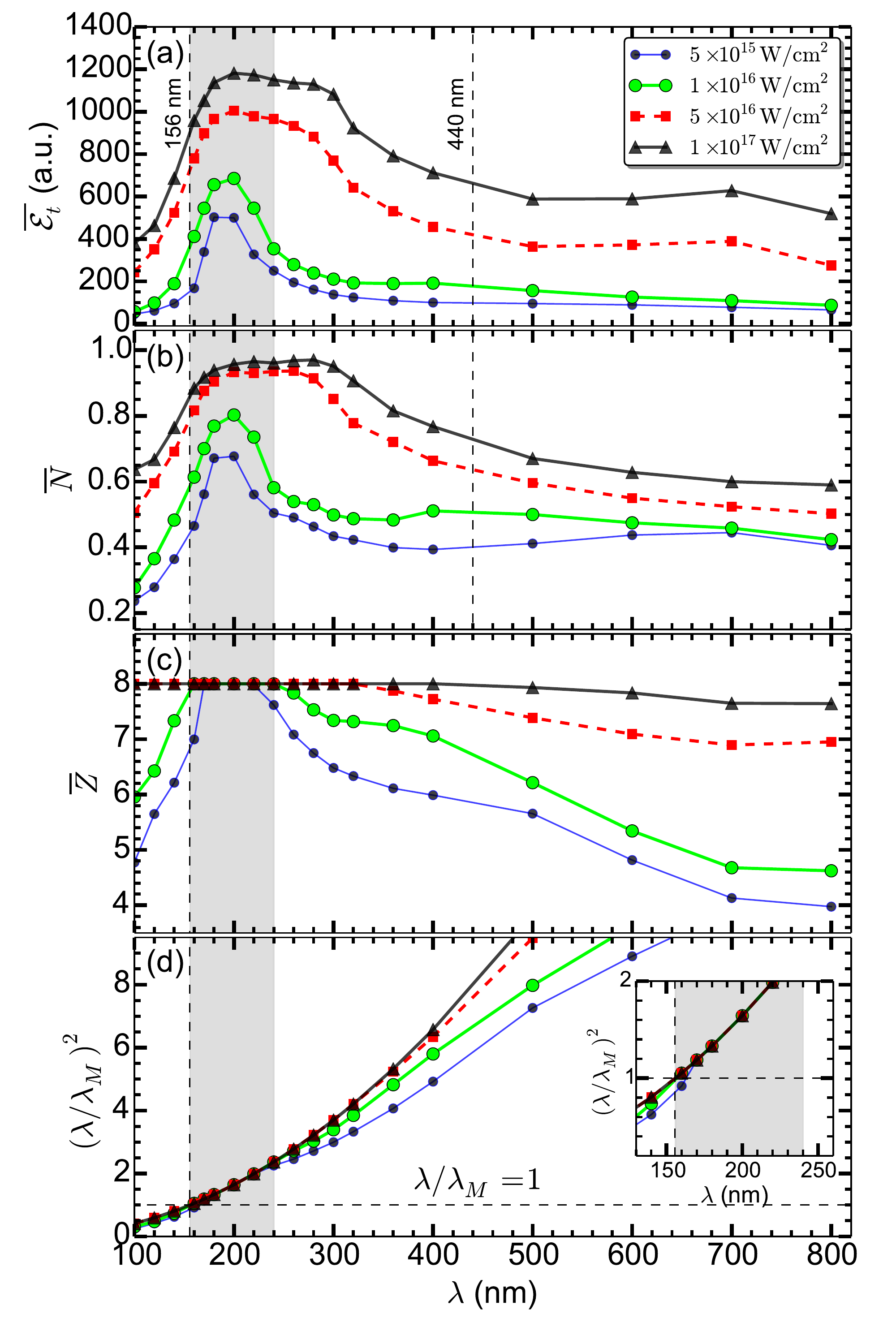}
\caption{(color online) 
Average total absorbed energy $\overline{\mathcal{E}}_t$ per atom~(a), corresponding fractional outer 
ionization $\overline{N}$ of electrons~(b), average  $\overline{Z}$ of argon ions~(c), and 
$(\lambda/\lambdaM)^2$~(d) versus $\lambda$ for the same argon cluster 
($R_0=2.91$~nm, $N=1791$) at the end of 13.5~fs continuous 
pulses $E_l(t) = E_0\sin(\omega t)$ 
(with no uncertainty in $\omega$) of same 
$I_0 \approx 5\times 10^{15}, 10^{16}, 5\times 10^{16}, 10^{17}\,\Wcmcm$ as in Fig.\ref{fig3abcd}. 
For a given $I_0$, continuous pulses of different $\lambda$ are chosen by keeping pulse duration 
$\tau \approx 13.5$~fs as constant. Vertical dashed lines indicate $\lambdaM$ where 
absorption maxima are strictly expected by LR for respective $\overline{Z} = 1,8$. Shaded bar highlights 
that absorption maxima are red-shifted in the marginally over-dense regime of 
$\lambda/\lambdaM\approx 1-1.5$, which is more clear in the inset of (d). Clearly, redshifts in 
absorption peaks are not due to uncertainties in frequencies of shorter pulses.
}
\label{fig5abcd}
\end{figure}

One may surmise that whether redshifts of absorption peaks in Fig.\ref{fig3abcd} are due to the frequency broadening of short 
pulses used here. Note that pulse duration is 
fixed at $\tau \approx 13.5$~fs which corresponds $n\!=\!5$ laser cycles at $\lambda =800$~nm and $n=40$ laser cycles 
at $\lambda =100$~nm. So the uncertainties $\Delta\omega=\omega/n$ in frequencies $\omega_{2,3}=(1 \pm 1/n)\omega$ 
[see below Eq.(\ref{eq:laserfield}) for $\omega_{2,3}$] are greatly reduced from $0.2\omega$ to $0.025\omega$ as $\lambda$ is changed 
from 800~nm to 100~nm. 
Corresponding to peak values of $\overline{Z}\approx 7-8$ in Fig.\ref{fig3abcd}(c), for intensities $I_0 \approx 5\times 10^{15} - 10^{17}\,\Wcmcm$ near the absorption peak, Mie-resonance wavelengths lie in the range of $\lambdaM \approx 166-156$~nm. The uncertainties in frequencies at these laser wavelengths $\lambda$ are $\Delta\omega\approx (800\times 5/166)^{-1}\omega\approx 0.0415\omega$ and $\Delta\omega\approx (800\times 5/156)^{-1}\omega\approx 0.04\omega$ where $(800\times 5/\lambda)$ is the $n$ for $\lambda$; and corresponding redshifts of 
$\Delta\lambda\approx 0.0415\times 166\approx 7$~nm and $\Delta\lambda\approx 0.04\times 156\approx 6$~nm may be expected in absorption peaks. However, if we see MD results in Fig.\ref{fig3abcd}, absorption peaks occur near $\lambda=\Lambda_d\approx 200$~nm and estimated redshifts of $\Lambda_d$ from $\lambdaM$ range from $\Delta\Lambda_d\approx 30$~nm to $\Delta\Lambda_d\approx 45$~nm (depending on the laser intensity).
The expected redshifts of $\Delta\lambda \approx 6-7$~nm (obtained above) 
due to the uncertainties in laser frequencies of shorter pulses 
are significantly small and insufficient to explain estimated redshifts 
$\Delta\Lambda_d\approx 30-45$~nm of $\Lambda_d$ form respective $\lambdaM=166-156$~nm.

To establish the argument that redshifts of absorption peaks 
in Fig.\ref{fig3abcd} (also in Figs.\ref{fig1abc},\ref{fig2ab} in the model), are not {\textcolor{blue}{only}} due to the frequency broadening of short pulses, we 
now simulate the argon cluster of Fig.\ref{fig3abcd} with continuous laser pulses $E_l(t)=E_0\sin(\omega t)$ (having no uncertainty in $\omega$) 
of same $I_0 \approx 5\times 10^{15}, 10^{16}, 5\times 10^{16}, 10^{17}\,\Wcmcm$ and $\lambda=100-800$~nm 
as in Fig.\ref{fig3abcd}, and stop the pulses after $\tau =13.5$~fs (for 5-cycles at 800~nm). In this case, pulse-envelops may 
be regarded as rectangular, each having a duration of $\tau=13.5$~fs. 
Corresponding results are displayed in Figs.\ref{fig5abcd}(a)-\ref{fig5abcd}(d) which bear 
same meaning of respective Figs.\ref{fig3abcd}(a)-\ref{fig3abcd}(d) (see also caption of 
Fig.\ref{fig5abcd}). Compared to Fig.\ref{fig3abcd}, for a given 
$I_0$ and $\lambda$, absorbed energy $\overline{\mathcal{E}}_t$ per atom, 
and fractional outer ionization $\overline{N}$ of electrons
systematically increase (average $\overline{Z}$ also systematically increase except near the closed
shell of argon atoms, where $\overline{Z}=8$ is saturated)  due to more energy carried by a continuous pulse than a
$\sin^2$-enveloped pulse (\ref{eq:laserfield}) (used in Fig.\ref{fig3abcd}) of same duration $\tau=13.5$~fs. Even, at the 
lowest intensity $I_0 \approx 5\times 10^{15}\,\Wcmcm$, average $\overline{Z}$ now reaches a saturation at
$\overline{Z}=8$ within the band of $\lambda\approx 156-240$~nm. 
For a given $I_0$, however, redshifts of peaks of $\overline{\mathcal{E}}_t$ 
and $\overline{N}$ in Figs.\ref{fig5abcd}(a)-\ref{fig5abcd}(b) are either nearly {\textcolor{blue}{the}} same as those of Figs.\ref{fig3abcd}(a)-\ref{fig3abcd}(b) or have the 
tendency to be more red-shifted from the static Mie-resonance wavelength $\lambdaM=156$~nm (dashed vertical line). 
Nevertheless, redshifts of absorption peaks also occur with continuous pulses (in Fig.\ref{fig5abcd}) 
and the uncertainty in the frequency broadening of a shorter pulse of $\tau=13.5$~fs is insufficient to explain the 
estimated redshift of $\Lambda_d$ from $\lambdaM$ in this work. 

\vspace{-0.6cm}
\subsubsection{Resonance shift with cluster size variation}
\begin{figure}
\includegraphics[width=1.0\linewidth]{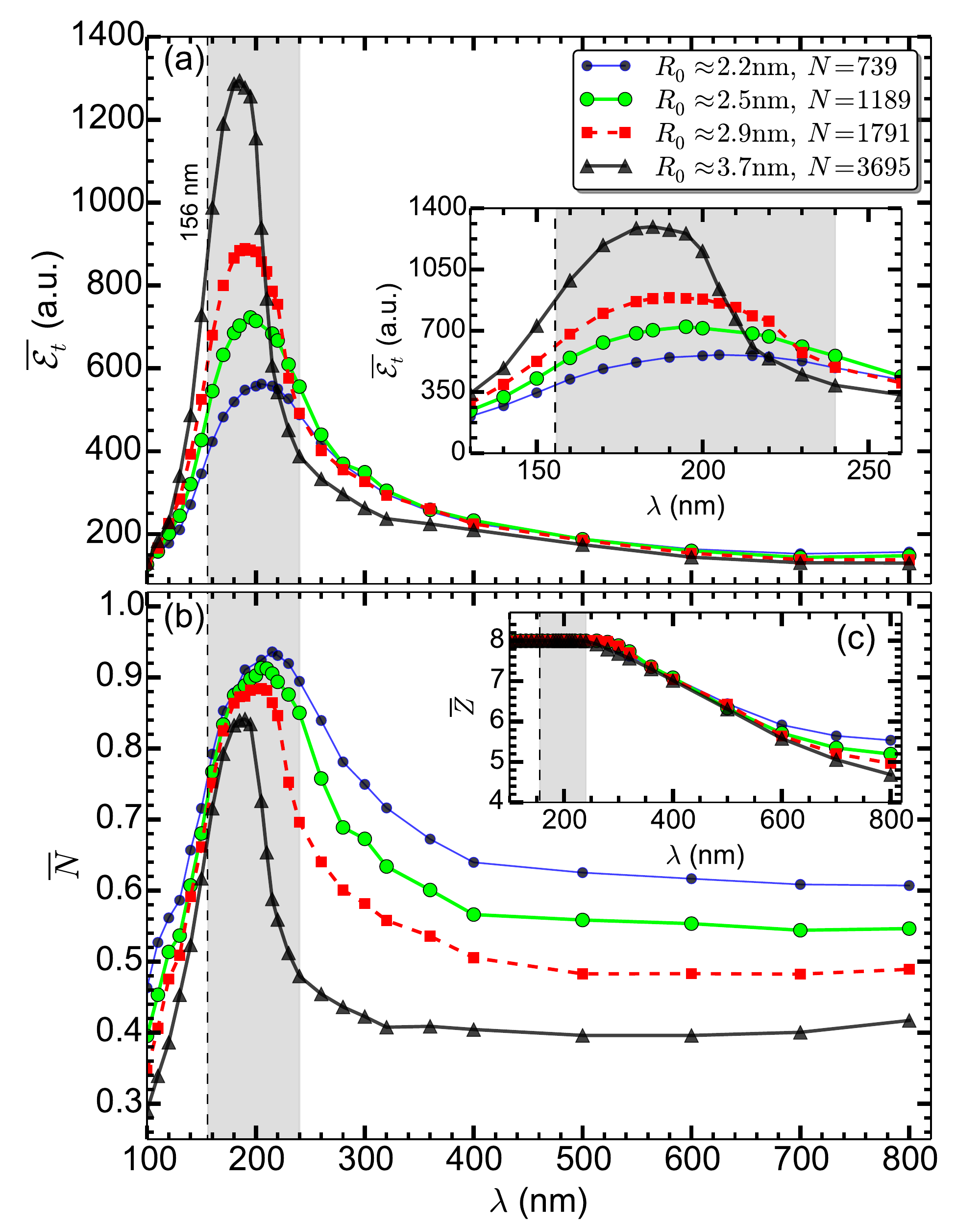} 
\caption{(color online) 
Average total absorbed energy $\overline{\mathcal{E}}_t$ per atom~(a), fractional outer 
ionization $\overline{N}$ of electrons~(b) and average  $\overline{Z}$ of argon ions~(c) versus $\lambda$ 
for different argon clusters of radii $R_0\approx 2.2, 2.5, 2.9, 3.7$~nm
with respective number of atoms $N=739, 1189, 1791, 3695$ (including cluster of Fig.\ref{fig3abcd}) 
after the laser pulse of fixed $I_0 \approx 5\times 10^{16}\,\Wcmcm$ as in Fig.\ref{fig3abcd}
and fixed $\tau \approx 13.5$~fs.
Other conditions are same as Fig.\ref{fig3abcd}.
Vertical dashed line indicate $\lambdaM$ where absorption peak is expected by LR for $\overline{Z} = 8$. Shaded bar highlights that absorption maxima are red-shifted in 
the marginally over-dense regime of $\lambda/\lambdaM\approx 1-1.5$ [more clear 
in the inset of (a)]. Redshift in absorption peak decreases as cluster size increases.
}
\label{fig6ab}
\end{figure}
%
So far it is not known how the redshift of the absorption peak changes with the cluster size variation. 
For completeness, we now study argon clusters of four different sizes of $R_0\approx 2.2, 2.5, 2.9, 3.7$~nm
(including the cluster of Fig.\ref{fig3abcd}) with respective number of atoms $N=739, 1189, 1791, 3695$  
while keeping the peak intensity fixed at $I_0 \approx 5\times 10^{16}\,\Wcmcm$ and the same laser 
pulse conditions of Fig.\ref{fig3abcd} with $\tau=13.5$~fs. 

As in Figs.\ref{fig3abcd}(a)-\ref{fig3abcd}(c), $\overline{\mathcal{E}}_t$, 
$\overline{N}$ and $\overline{Z}$ versus 
$\lambda$ for different $R_0$ are shown in respective Figs.~\ref{fig6ab}(a)-\ref{fig6ab}(c) where different $\overline{Z}$ do not differ 
much for $\lambda<500$~nm and saturate at $\overline{Z}=8$ irrespective of cluster sizes in the 
marginally over-dense band of $\lambda = 156-240$~nm.
The maximum number of electrons created in the largest and the smallest cluster are respectively 
$\Ne\approx N\overline{Z} = 29560, 5912$, corresponding maximum outer ionized electrons are 
$\Ne^{om}\approx 24535,5498$ with $\overline{N}\approx 0.83, 0.93$ [from Fig.\ref{fig6ab}(b)], and 
number of outer ionized electrons per atom can be estimated to be $\Ne^{om}/N\approx 6.64,7.44$ respectively. 
It means that a bigger cluster retains more electrons per atom by exerting relatively higher restoring 
force than a smaller cluster for the same laser intensity, although absolute outer ionization of electrons 
from a bigger cluster may exceed a small cluster. 
Thus, for the same $I_0$ and $\tau$, as the cluster size increases, the higher 
restoring of ions on the electrons in a bigger cluster reduces $\overline{N}$, reduces the 
redshift of the peak of $\overline{\mathcal{E}}_t$ and $\overline{N}$ towards $\lambdaM=156$~nm (dashed vertical line),
{\em but} redshift {\em still} persists in the marginally over-dense band of $\Lambda_d=(1-1.5)\lambdaM$. 

{\textcolor{blue}{
\begin{figure}
{\textcolor{blue}{
\includegraphics[width=0.8\linewidth]{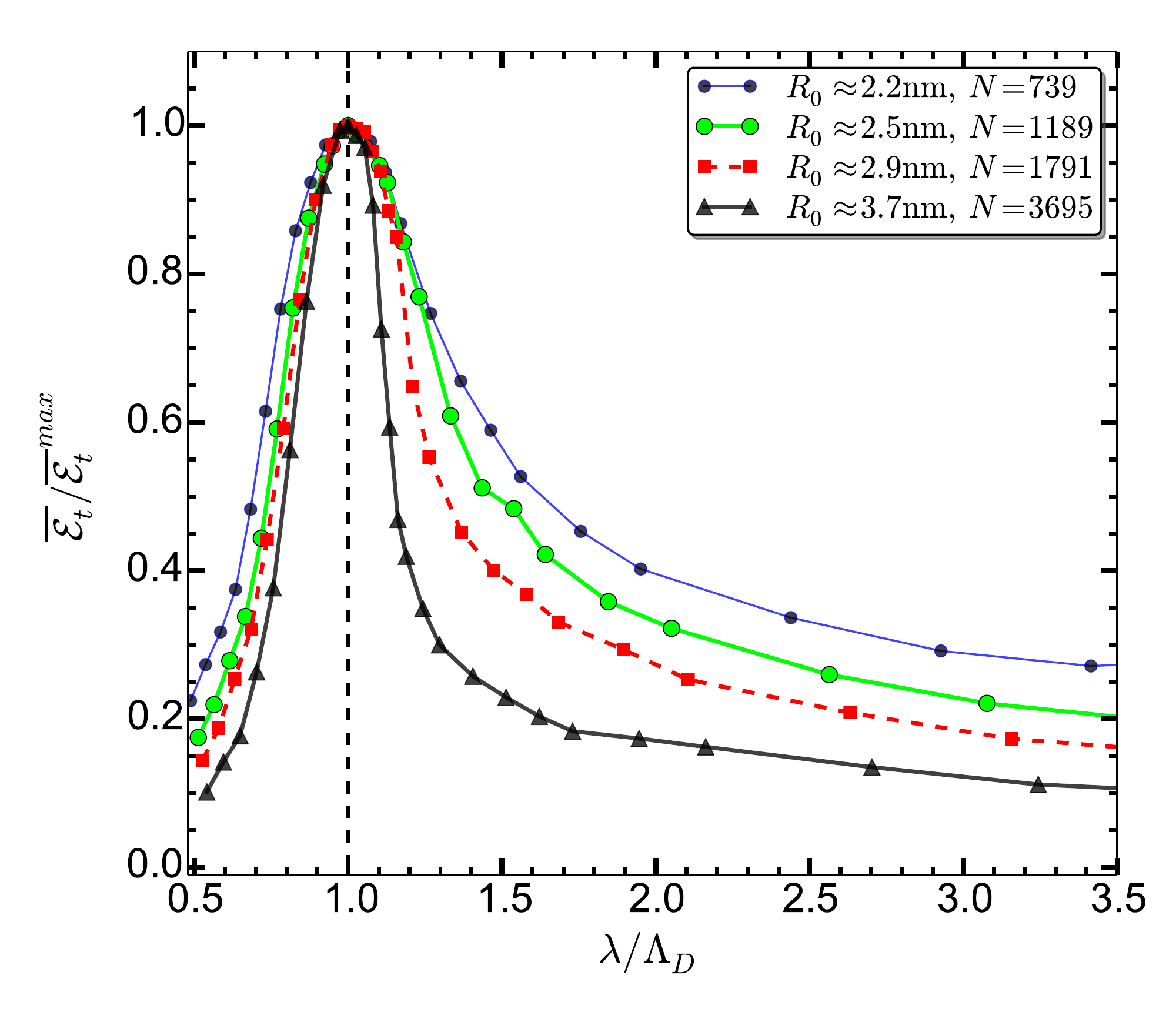} 
\caption{(color online) %
Normalized absorbed energy $\overline{\cal{E}}_t/\overline{\cal{E}}_t^{max}$ versus $\lambda/\Lambda_D$ corresponding to Fig.6 where $\overline{\cal{E}}_t^{max}=\max(\overline{\cal{E}}_t)$ at the corresponding shifted wavelength $\Lambda_D$ for each $R_0$.
}
\label{fig7}
}}
\end{figure}
One may vaguely imagine for a universal absorption curve. From Fig.\ref{fig6ab} we note maximum absorption $\overline{\cal{E}}_t^{max} = \max(\overline{\cal{E}}_t)$ and the respective shifted wavelength $\Lambda_D$ for each $R_0$. Figure~\ref{fig7} shows corresponding normalized energy $\overline{\cal{E}}_t/\overline{\cal{E}}_t^{max}$ versus $\lambda/\Lambda_D$. 
It is found that there is neither a universal curve, nor whole curves themselves behave 
the same for all cluster sizes. A similar observation is also confirmed for Fig.\ref{fig3abcd} at 
different peak intensities.
}}


\section{Summary and Conclusion}\label{sec6}
We study interaction of short 5-fs (fwhm) laser pulses with an argon cluster using MD simulation and RSM to find out the wavelength regime where laser absorption in argon cluster is maximized for a given intensity and pulse energy. It seems trivial for many researchers (in the field of laser-plasma interaction) to answer immediately that maximum absorption should occur at the wavelength $\lambdaM$ of linear resonance (LR) irrespective of laser intensity, as in the nano-plasma model \cite{Ditmire_PRA53} or in the collective oscillation model.
However we find that, for a given laser pulse energy and a cluster, at each peak intensity there exists a $\lambda$ -- shifted from the expected LR wavelength of $\lambdaM$ -- that corresponds to a {\em unified dynamical} LR (coined as UDLR) at which evolution of the cluster happens through very efficient unification of possible resonances in various stages, including (i) the LR in the initial time of plasma creation, (ii) the LR in the Coulomb expanding phase in the later time and (iii) anharmonic resonance in the marginally over-dense regime for a relatively longer pulse duration below 5-fs (fwhm), leading to maximum laser absorption accompanied by maximum removal of electrons from cluster and also maximum allowed average charge states $\overline{Z}$ of argon atoms in the cluster. 
Increasing the laser intensity, the absorption maxima is found to shift
to a higher wavelength in the band of $\Lambda_d\approx (1-1.5)\lambdaM$
than staying permanently at the expected $\lambdaM$, e.g., $\lambdaM\approx 156$~nm for $\overline{Z}=8$ [see Fig.\ref{fig3abcd}]. 
The simple RSM also corroborates the wavelength shift of the absorption peak (in spite of the absence of cluster charging and expansion) as found in MD. Thus RSM and MD un-equivocally prove that maximum absorption in a laser driven cluster happens at a shifted $\lambda$ in the marginally over-dense regime of $\lambda\approx (1-1.5)\lambdaM$ instead of $\lambdaM$ of LR
for all intensities $10^{15}\,\Wcmcm-5\times 10^{17}\,\Wcmcm$. 
Therefore, if an experiment or a simulation is performed at the static LR wavelength which is {\em often} decided {\em a priori}, the absorption will not be maximum and may lead to a conclusion that resonance has no role \cite{Petrov2005_PRE,Petrov2006} for laser absorption. Instead, at a given intensity, an efficient unification of resonances (i.e., UDLR) happens at a shifted wavelength in the band of $\Lambda_d\approx (1-1.5)\lambdaM$ that leads to maximum laser absorption. Our results may be useful to guide an optimal condition experiment with argon cluster in the short pulse regime where maximum conversion of energy from laser to particles is required.

{\textcolor{blue}{It may be questioned that laser absorption is affected due to carrier envelope phase (CEP) effects of short 13.5 fs pulses and cosequently the red-shift of the absorption peak as reported here may differ for different CEPs. We have confirmed by MD simulations that the effect of CEP is not appreciable for absorbed energy and outer ionization for the laser pulse duration of 13.5~fs for all wavelengths $100-800$~nm and intensities $<5 \times10^{17}\,\Wcmcm$; and the red-shift of an absorption peak is hardly affected by different CEPs.}}

\nocite{*}
\bibliography{ArgonCluster_WavelengthDepSMMKv2}
\end{document}